\documentclass[a4paper,fleqn,usenatbib]{mnras}

\usepackage{newtxtext}
\usepackage{newtxmath}

\usepackage[T1]{fontenc}
\usepackage{ae,aecompl}

\usepackage{graphicx}	
\usepackage{amsmath}	
\usepackage{amssymb}	
\usepackage{times}
\usepackage{hyperref}
\usepackage[normalem]{ulem}

\usepackage{titlesec} 

\titleformat{\subsection}[runin]
  {\normalfont\normalsize\bfseries}{\thesubsection}{1em}{}
\titleformat{\subsubsection}[runin]
  {\normalfont\normalsize\bfseries}{\thesubsubsection}{1em}{}

\titlespacing*{\subsection}{0pt}{.7\baselineskip}{\baselineskip}

\newcommand{\voles}{{\sc vole}s}

\newcommand{\Rv}{\ensuremath{ R_{\rm v} }}   
\newcommand{\hMpc}{\ensuremath{ ~h^{-1} \rm{Mpc} }}	

\usepackage{color}
\definecolor{colorChanges}{rgb}{.8,.2,1.}
\definecolor{colorDeletes}{rgb}{1.,.2,.2}

\title[Voles]{Weak lensing by voids in weak lensing maps}

\author[Davies et al.]{
Christopher T.~Davies$^{1}$\thanks{E-mail: christopher.t.davies@durham.ac.uk},
Marius Cautun$^{1}$
and Baojiu Li$^{1}$
\\
$^{1}$Institute for Computational Cosmology, Department of Physics, Durham University, South Road, Durham DH1 3LE, UK\\
}

\pubyear{2018}

\begin{document}
\label{firstpage}
\pagerange{\pageref{firstpage}--\pageref{lastpage}}
\maketitle

\begin{abstract}
Cosmic voids are an important probe of large-scale structure that can constrain cosmological parameters and test cosmological models. 
We present a new paradigm for void studies: void detection in weak lensing convergence maps. 
This approach identifies objects that relate directly to our theoretical understanding of voids as underdensities in the total matter field and presents several advantages compared to the customary method of finding voids in the galaxy distribution.
We exemplify this approach by identifying voids using the weak lensing peaks as tracers of the large-scale structure.
We find self-similarity in the void abundance across a range of peak signal-to-noise selection thresholds.
The voids obtained via this approach give a tangential shear signal up to $\sim40$ times larger than voids identified in the galaxy distribution.
\end{abstract}




\section{Introduction}
\label{sect:intro}

The large-scale structures of the Universe, collectively called the cosmic web, describe the  matter distribution in our Universe in the forms of structures such as voids, sheets, filaments and knots. These structures result from the anisotropic gravitational collapse of matter on cosmic scales. These components are intertwined in a complex web where the knots form at the intersections of filaments, filaments form at the intersections of sheets, and voids occupy the underdense space between all three \citep{LSS}.

Voids, which represent large regions mostly devoid of matter and galaxies, have attracted a lot of interest as powerful probes of cosmological parameters \citep{Lavaux2012,Hamaus2016}, dark energy \citep{Li2011,Bos2012,Pisani2015} and dark matter \citep{Massara2015}. Voids are especially useful for testing cosmological models that make environmentally dependent predictions, such as the fifth force of modified gravity theories which, while screened in high density regions, attains maximum values in voids \citep{Clampitt2013,Cai2015,modified-potential,Falck2018,Baker:2018mnu}. The largest constraining power of voids comes from measuring their total matter content \citep[e.g.][]{SHED}, which can be achieved via gravitational lensing \citep{Melchior:2014, Void-Obs-SDSS,Gruen:2015jhr,DES-voids,CMB-voids}, redshift space distortions \citep{Hamaus2015,RSD} as well as the integrated Sachs-Wolfe effect in the cosmic microwave background \citep{Granett2008,Nadathur2016}. In particular, weak lensing (WL) measurements using upcoming large area and deep imaging surveys such as \textsc{euclid} and \textsc{lsst} will result in tight constraints on the mass profile of voids \citep{Krause2013,SHED}.

Theoretically, voids correspond to low density regions in the large-scale matter field \citep{Sheth2004,voids-review,Aragon-Calvo2013}. However, because the full mass distribution is not easily observable, observational studies typically identify voids using the galaxy distribution \citep[e.g.][]{Nadathur2016}. Due to the sparsity of galaxy tracers and their bias, which depends on environment \citep{Neyrinck2014}, the relation between matter and galaxy voids is a complex one, with galaxy voids being typically less underdense than would have otherwise been identified using the full matter density field. This could potentially weaken the lensing signals (which are produced by the total matter) from galaxy voids, and, due to difficulties in simulating galaxies in cosmological volumes, it is also more challenging to test cosmology using galaxy void properties such as abundances and sizes \citep[see, e.g.,][]{SHED}.

In this work, we propose a new paradigm for void studies: the identification of voids from weak lensing convergence maps, which we refer to as \voles{} (VOids from LEnsing). The convergence field represents the projected line-of-sight density field weighted by the lensing kernel, and thus the identified underdensities correspond to voids in the projected density field (for example \citealt{Chang2018} have shown that the deepest minima of the convergence field have good correspondence to galaxy voids). This approach represents a simple way of finding voids that relates directly to our theoretical understanding of voids as underdensities in the total matter field. As void lensing is a key observable, it is only natural to identify voids and extract their lensing signal from the very same observations, such as a convergence map. \voles{} not only help avoid some of the disadvantages of galaxy voids, but also allow for a more complete exploitation of lensing maps by naturally combining \voles{} with other statistics, such as WL peaks and Minkowski functionals.

There are many void finders in the literature \citep[e.g. see the void comparison project of][]{Colberg2008} and here we choose to illustrate our methodology using the tunnel finding method \citep{SHED}, but, in principle, many of the previous void finding approaches can be applied to the lensing convergence maps by using the convergence field itself, rather than using peaks as tracers. Our choice of tunnels is based on the \citet{SHED} and \citet{Paillas2018} studies which find that WL by tunnels identified in the galaxy distribution is the most promising method for testing a wide range of modified gravity theories.

\section{Method}

\subsection{N-body and ray-tracing simulations:}
\label{subsect:sims}

The WL maps used in this analysis are made using an analytical on-the-fly ray tracing algorithm, {\sc ray}-{\sc ramses} \citep{on-the-fly,Ray-Ramses}, which is inbuilt in the publicly available N-body and hydrodynamical adaptive mesh refinement simulation code {\sc ramses} \citep{Ramses}. The N-body evolution part is done using the default RAMSES code. To construct the first map, five independent realisations of simulations evolving $1024^3$ particles in a $512\hMpc$ box are tiled together to form a light cone up to a source redshift $z_s=1$ (see Fig.1 of \citealt{Ray-Ramses} for an illustration). The second map is constructed from the same box repeated 5 times. The cosmological parameters adopted are $\Omega_m = 0.32$, $\Omega_{\Lambda} = 0.68$ and $H_0 = 67~\rm{km~s^{-1}~Mpc^{-1}}$. 
The two WL convergence maps cover a field of view of $10 \times 10$ deg$^2$ with a resolution of $2048^2$ pixels. In order to use information from separate maps in conjunction with each other, we respectively subtract the mean convergence value of each map, to give us zero-mean convergence maps.

\subsection{Galaxy shape noise:}
\label{subsect:noise}
The convergence field is determined observationally by averaging over a large number of source galaxies, which, due to their intrinsic ellipticity, leads to measurement uncertainties. This effect is known as galaxy shape noise (GSN) and can be a main uncertainty source on small angular scales.
To allow our method to be interpreted in the context of observations, we generate WL maps with added GSN and compare the \voles{} identified with and without GSN. 
For each pixel of the WL maps, we add GSN by drawing from a Gaussian with standard deviation,
\begin{equation}
\sigma_{\rm pix}^{2} =  \sigma_{\rm int}^{2} \,/\, 2 \theta_{\rm pix}^{2} n_{\rm gal}
\label{EQ:} \;,
\end{equation}
where $\sigma_{\rm int}$ is the dispersion of the source galaxy intrinsic ellipticity, $\theta_{\rm pix}$ the angular width of each pixel, and $n_{\rm gal}$ the number density of source galaxies.  Here we use $\sigma_{\rm int} =0.4$ and $n_{\rm gal} = 40$ arcmin$^{-2}$ corresponding to \textsc{lsst} \citep{LSST2009}.

\subsection{Peak extraction:}

In a first step, we identify peaks in the convergence map. In order to suppress GSN, we smooth the convergence map using a Gaussian window with a smoothing scale, $\theta_s$, of 2.5 arcmin (unless otherwise stated). We define a peak as a pixel whose convergence value is greater than that of its 8 neighbours. We also trim the peaks in each map within one smoothing length of the edge of the map to exclude the boundary effects of smoothing a finite map. Each peak is characterised by the lensing convergence at its position, which we express as a signal-noise-ratio (SNR), $\nu\equiv\kappa/\sigma$, where $\sigma$ is the standard deviation of the smoothed convergence maps, which is $0.011$ and $0.012$ for the maps without and with GSN, respectively.

\begin{figure}
	\centering
	\includegraphics[width=\columnwidth]{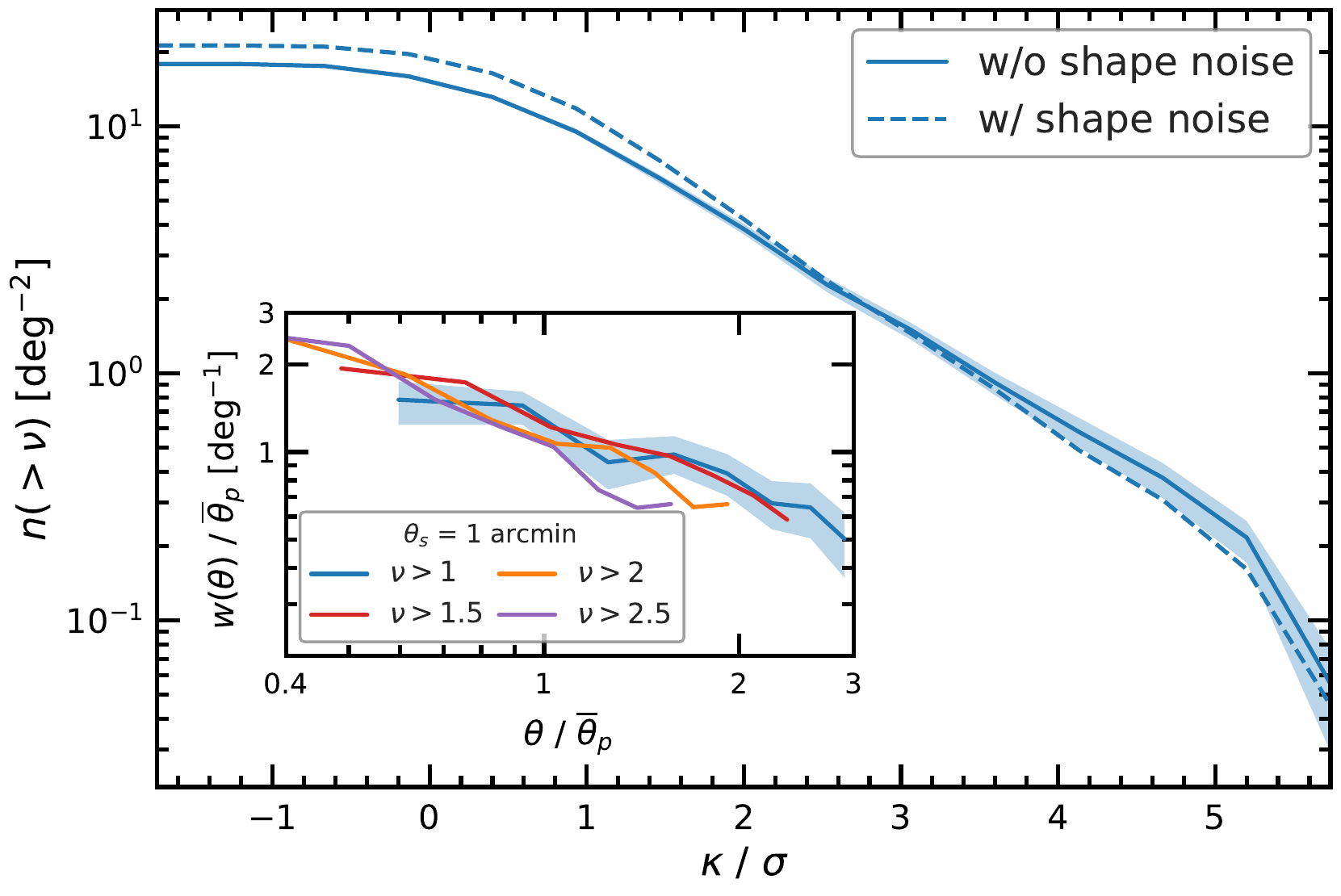}
    \vskip -.3cm
	\caption{(Colour Online) Cumulative number per unit area of convergence peaks as a function of peak SNR, $\nu=\kappa/\sigma$, for the maps with and without GSN. 
The inlay plot shows the two-point peak correlation functions, $w(\theta)$, for peak catalogues with $\nu$ thresholds indicated in the legend, identified using $\theta_s = 1$ arcmin, with the peak pair separation $\theta$ (horizontal axis) and $w(\theta)$ (vertical axis) both scaled by $\bar{\theta}_p$, the  mean peak separation in the respective peak catalogue. Though noisy due to the small map size, the rescaled curves show self-similarity for a range of $\nu$ thresholds. The shaded regions show 1$\sigma$ uncertainties.}
	\label{fig:peak distribution}
\end{figure}

\begin{figure*}
	\centering
	\includegraphics[scale=0.6]{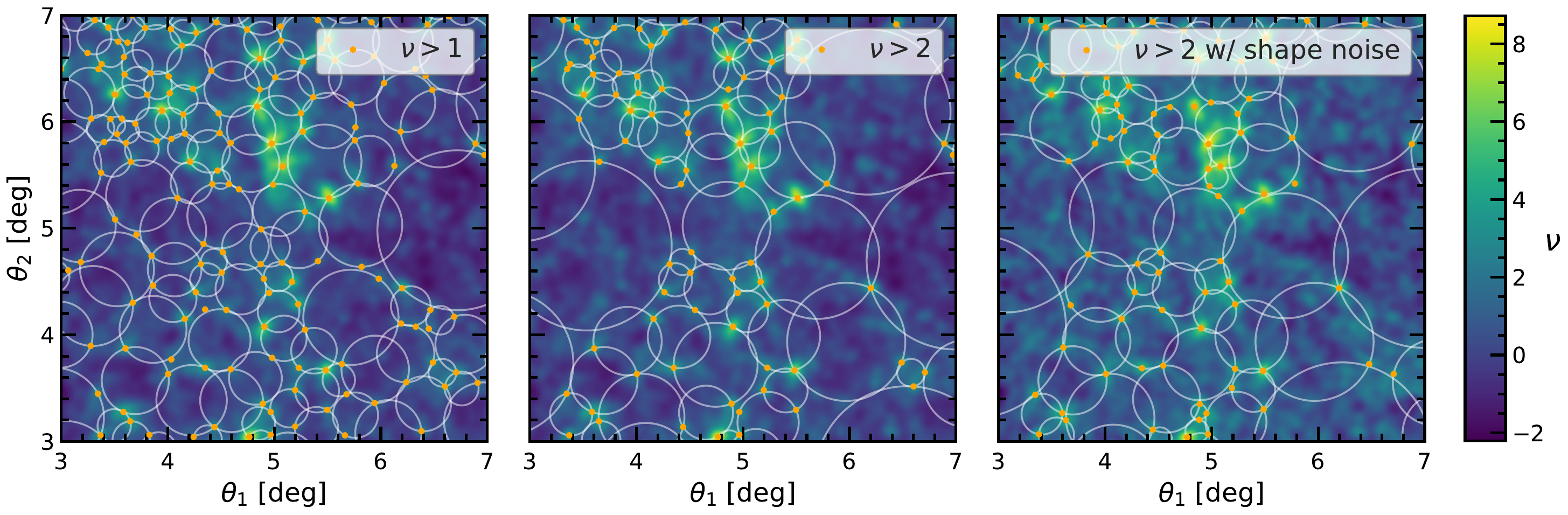}
    \vskip -.1cm
	\caption{(Colour Online) Visualisation of 2D voids found in a small region of the smoothed lensing convergence maps (with and without galaxy shape noise). The panels show the voids for two convergence peak selection criteria, from left to right: $\nu >1$, $\nu >2$, and the third panel shows $\nu >2$ with galaxy shape noise added to the map. The background colours show the convergence map, expressed in terms of $\nu$. The dots show the peaks that satisfy each selection criterion, and the circles show the voids identified in each peak catalogue. $\theta_1$ and $\theta_2$ denote angular coordinates of the map in two orthogonal directions.
    }
	\label{fig:void visualisation}
\end{figure*}

The resulting number density of WL peaks averaged across both of our convergence maps as a function of SNR, $\nu$, is shown in Fig.~\ref{fig:peak distribution}. It shows that the number of peaks is largest for small SNR, increasing from 2 $\rm{deg}^{-2}$ at SNR$\simeq3$ to 20 $\rm{deg}^{-2}$ at SNR$\simeq0$. Fig. \ref{fig:peak distribution} shows that our choice of $\theta_s$ (2.5 arcmin) leads to only small differences in the peak number densities. We have checked that both the number density and the two point correlation functions of peaks agree with previous studies \cite[e.g.][]{peak-dist-shirasaki,peak-dist-shan}. 
For each map, we generate peak catalogues by selecting all the peaks with SNR, $\nu$, above a given threshold value. Throughout this work, we mainly identify voids from three peak catalogues corresponding to $\nu >1$, $\nu >2$ and $\nu>3$, but in some places we also use catalogues with $\nu>1.5$ and $\nu>2.5$. 

\subsection{Void identification:}
\label{sect:Void generation}

We identify voids using the tunnel finding algorithm of \citet{SHED}, which is so-named because it has been developed to find regions in the projected distribution of galaxies that do not contain any galaxies.
The method can easily be extended and applied to the WL peak field by identifying the largest circles that are devoid of peaks. Thus, the tunnels correspond to circles in the 2D convergence map that contain no WL peaks. 

The tunnels are obtained by first constructing a Delaunay tessellation with its vertices chosen to be the WL peaks. By definition, the circumcircle of every Delaunay triangle does not enclose any WL peak. The WL peaks, which define the triangle, reside directly on its circumcircle. Thus, each circumcircle represents a candidate tunnel with radius, \Rv{}, and centre given by that of the corresponding circumcircle. We further discard any tunnels whose centres are found inside a larger tunnel. A visualisation of the tunnels found from the WL peak catalogues in one of our maps is shown in Fig.~\ref{fig:void visualisation}.

\subsection{Calculating void profiles:}

We calculate the convergence profiles of voids by using annuli of thickness $\Rv/N_{\rm bin}$, where $\Rv$ is the void radius and $N_{\rm bin}=20$, and then stack all the voids in terms of the scaled radial distance, $r/\Rv$. To get better statistics, we average over both lensing maps.

The tangential shear profile $\gamma_t(r)$ is calculated from the convergence profile using
\begin{equation}
	\gamma_t (r) = \bar{\kappa}(<r) - \kappa(r)  ; \quad \bar{\kappa}(<r) = \frac{1}{\pi r^2} \int_0^r 2 \pi r'\kappa (r'){\rm d}r'
	\label{Eq: shear_from_convergence} \;,
\end{equation}
where $\kappa(r)$ denotes the convergence at the radial distance $r$ and $\bar{\kappa}(<r)$ the mean enclosed convergence within $r$.

All of our uncertainties, including those of the convergence and shear profiles, are estimated using bootstrap sampling. For each of the two maps we generate 100 bootstrap resamples, which we then combine and quote the uncertainties obtained by taking the 16th and 84th percentiles of these resamples.

\section{Results}
\label{sect:results}

\begin{figure}
	\includegraphics[width=\columnwidth]{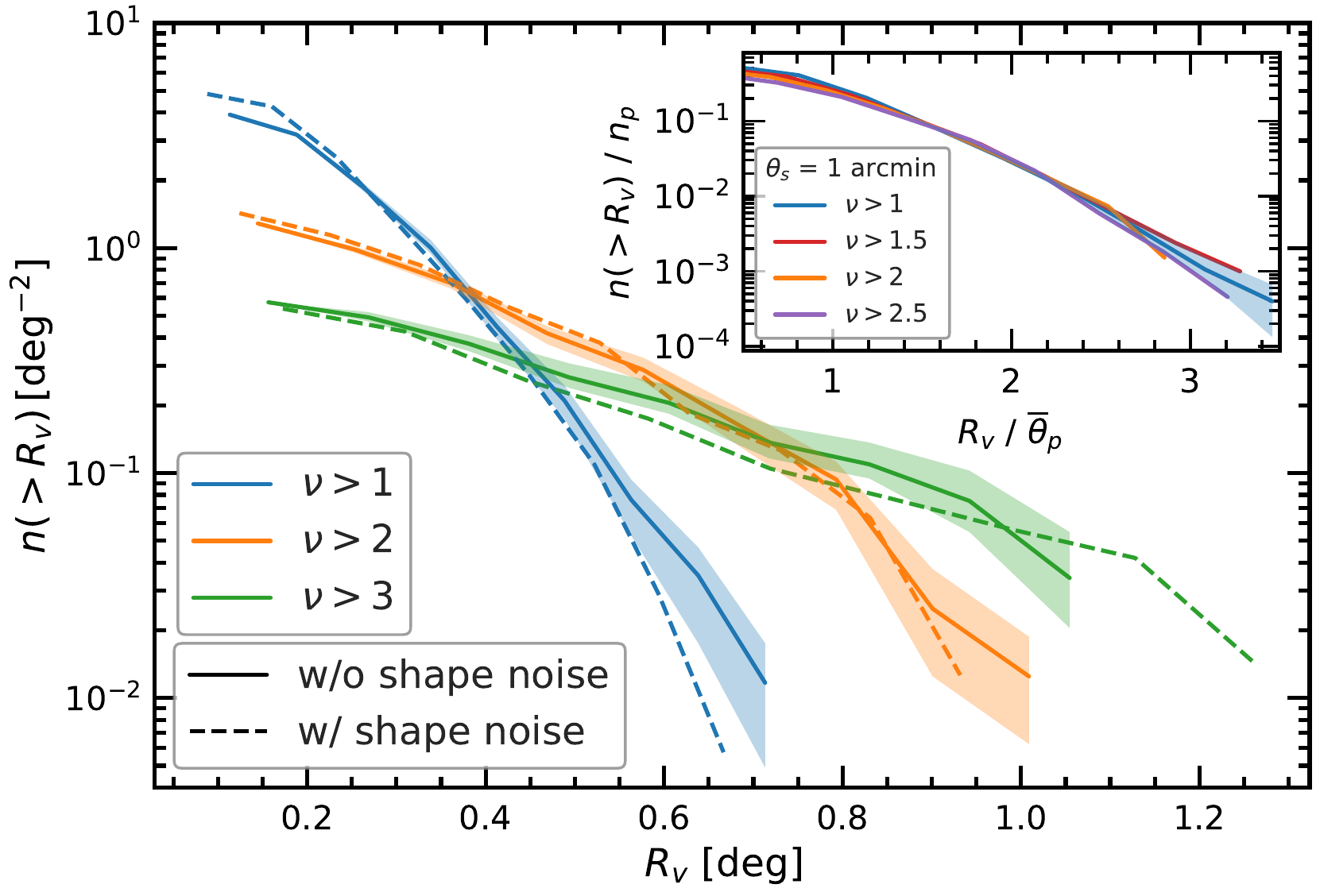}
    \vskip -.2cm
	\caption{(Colour Online) Cumulative number density of voids as a function of the void radius, $\Rv$. The curves correspond to void catalogues defined using WL peaks above different $\nu$ thresholds. 
{The inlay plot shows a universal relation in the void abundance for catalogues with $\theta_s$ = 1 arcmin, where $\Rv$, is scaled by the mean peak separation, $\bar{\theta}_p$, and $n(>\Rv)$, is divided by the mean peak number density, $n_p$, of the corresponding peak catalogue. The shaded regions show 1$\sigma$ uncertainties. }
}
	\label{fig:cumulative void radius distribution}
\end{figure}

\begin{figure}
	\centering
	\includegraphics[width=\columnwidth]{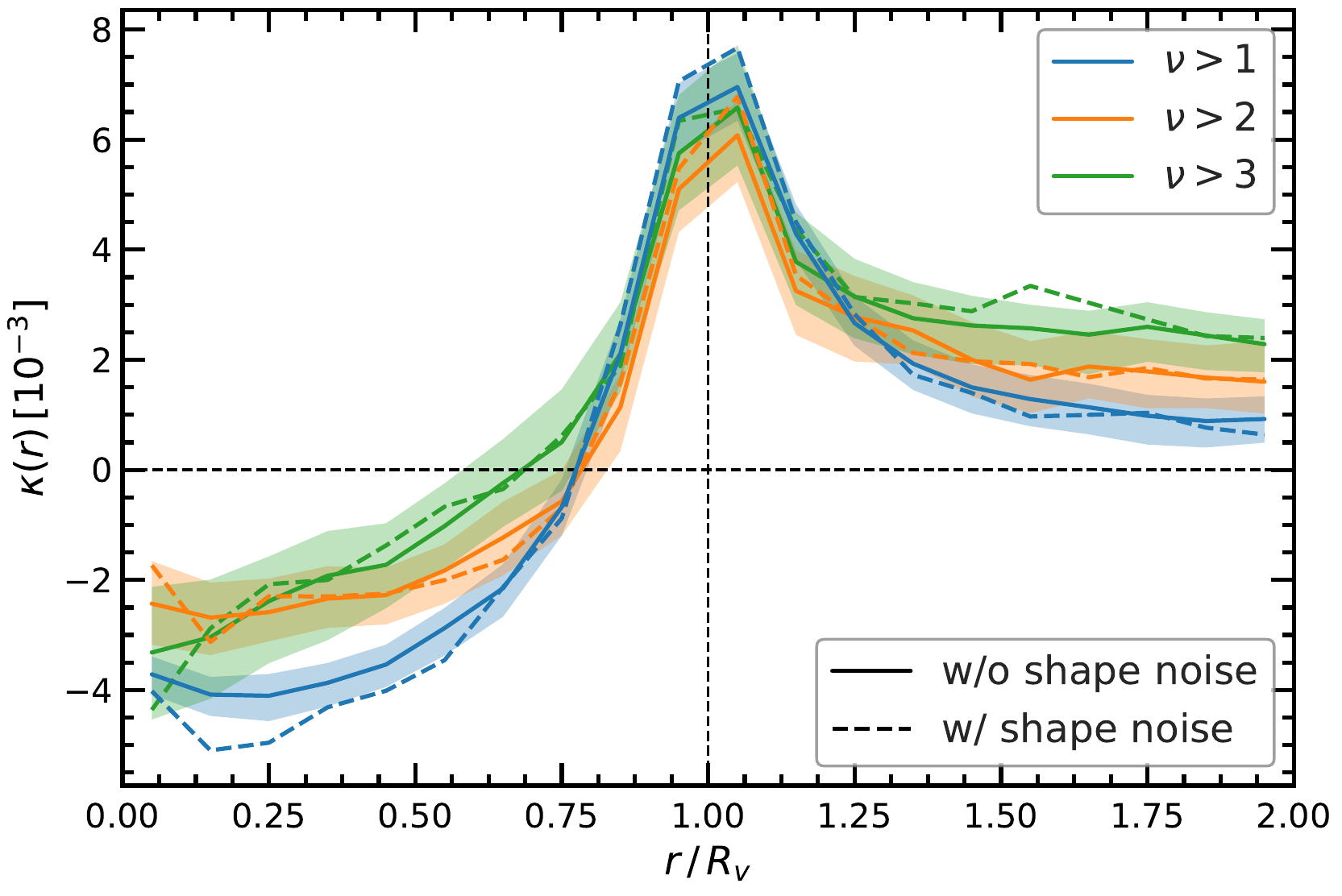}
    \vskip -.2cm
	\caption{(Colour Online) The stacked convergence profiles of WL voids. Each curve corresponds to an average over all of the objects in each of the void catalogues. The dashed lines show the added GSN case. The shaded regions show the 1$\sigma$ bootstrap uncertainties of the no GSN catalogues, which are similar to the GSN added case.} 
	\label{fig:Convergences}
\end{figure}

Fig.~\ref{fig:void visualisation} shows that, as the $\nu$ threshold increases, the WL peaks are more spread out, and less numerous, resulting in larger voids. This is as expected because high-SNR peaks generally correspond to more massive structures or more structures aligned along the line of sight, which are rarer \citep[e.g.,][]{Peak-origin}. The third panel shows how adding GSN can slightly alter the void catalogue. The dependence of void size on $\nu$ is quantitatively confirmed in the cumulative void radius distribution given in Fig.~\ref{fig:cumulative void radius distribution}, which shows that the $\nu >1$ void catalogue has more small voids and fewer large voids, with none above $\Rv\sim 0.7$ deg, while the $\nu >3$ catalogue has fewer small voids and more large voids with radii up to $\Rv\sim 1.1$ deg. Again we can see that GSN has a small impact on the \voles{} size distribution, where the addition of spurious peaks tend to slightly reduce the size of the \voles{}.
The inlay plot in Fig.~\ref{fig:cumulative void radius distribution} shows  self-similarity in the void abundance across the void catalogues with peak thresholds $\nu\geq1$, which is achieved by dividing the void radius by the mean peak separation and the void number density by the mean peak number density. We find this self-similarity for a range of smoothing scales, but we show only results for a small smoothing scale,  $\theta_s = 1$ arcmin,  which gives the largest number of \voles{} and thus provides the most stringent test of self-similarity. We also found that adding GSN has very little impact on the self-similarity of the void abundance, which allows us to choose the previously stated smaller smoothing scale. This self-similar scaled void abundance is likely related to the self-similarity in the peak two-point correlation functions from the peak catalogues with different SNR thresholds (see the inlay panel of Fig.~\ref{fig:peak distribution}), and we will present a detailed analysis on this, based on larger and more realistic lensing maps, in a forthcoming work.

\subsection{Void convergence profiles:}
Fig.~\ref{fig:Convergences} shows the convergence profiles as a function of scaled radial distance, $r/\Rv$, averaged over all voids in both lensing maps. The profiles are plotted up to twice the void radius to show how, at large distances, they return to background levels (which we have set to be 0). Each curve corresponds to one of the three void catalogues. 
For $r\lesssim0.75\Rv$, we find negative convergence values, which indicates that the voids are underdense in those regions. Interestingly, the $\nu>3$ void catalogue has the least underdense voids. This is to be expected, since the $\nu>3$ voids are the largest and can enclose inside them slightly overdense regions, i.e. with $\kappa>0$. As the peak SNR threshold used to identify voids decreases, the voids become smaller, they enclose fewer overdense regions and thus have lower overall $\kappa$ values. The maximum convergence is achieved at $r=\Rv$ and all three void catalogues have roughly similar maximum values. At even larger radii, the convergence profiles decrease towards the mean background value of $\kappa=0$. Of the three catalogues, $\nu>3$ voids take the longest to reach the background value, which is a manifestation of the fact that the large peak values that define the boundary of $\nu>3$ voids are typically found in large-scale overdense regions. Voids identified using high SNR peaks, i.e. $\nu\gtrsim2$, have the same profiles in WL maps with and without GSN, but differences appear when using low SNR peaks, i.e. $\nu\lesssim1$, where GSN can lead to spurious peaks, and thus spurious voids.

A final feature of the convergence profiles is that the width of the convergence maximum somewhat  decreases as we increase the $\nu$ threshold of the WL peaks used for finding voids. This is due to the profiles being plotted against the rescaled distance, $r/\Rv$, rather than the physical distance, $r$. The $\nu>2$ and $3$ voids actually have wider convergence maxima when expressed as a function of $r$, but this larger width is overcompensated by their even larger radii which results in a narrower maxima when expressed in rescaled distances.

\subsection{Tangential shear profiles:}
\label{subsect:gamma_profiles}

\begin{figure}
	\centering
	\includegraphics[width=\columnwidth]{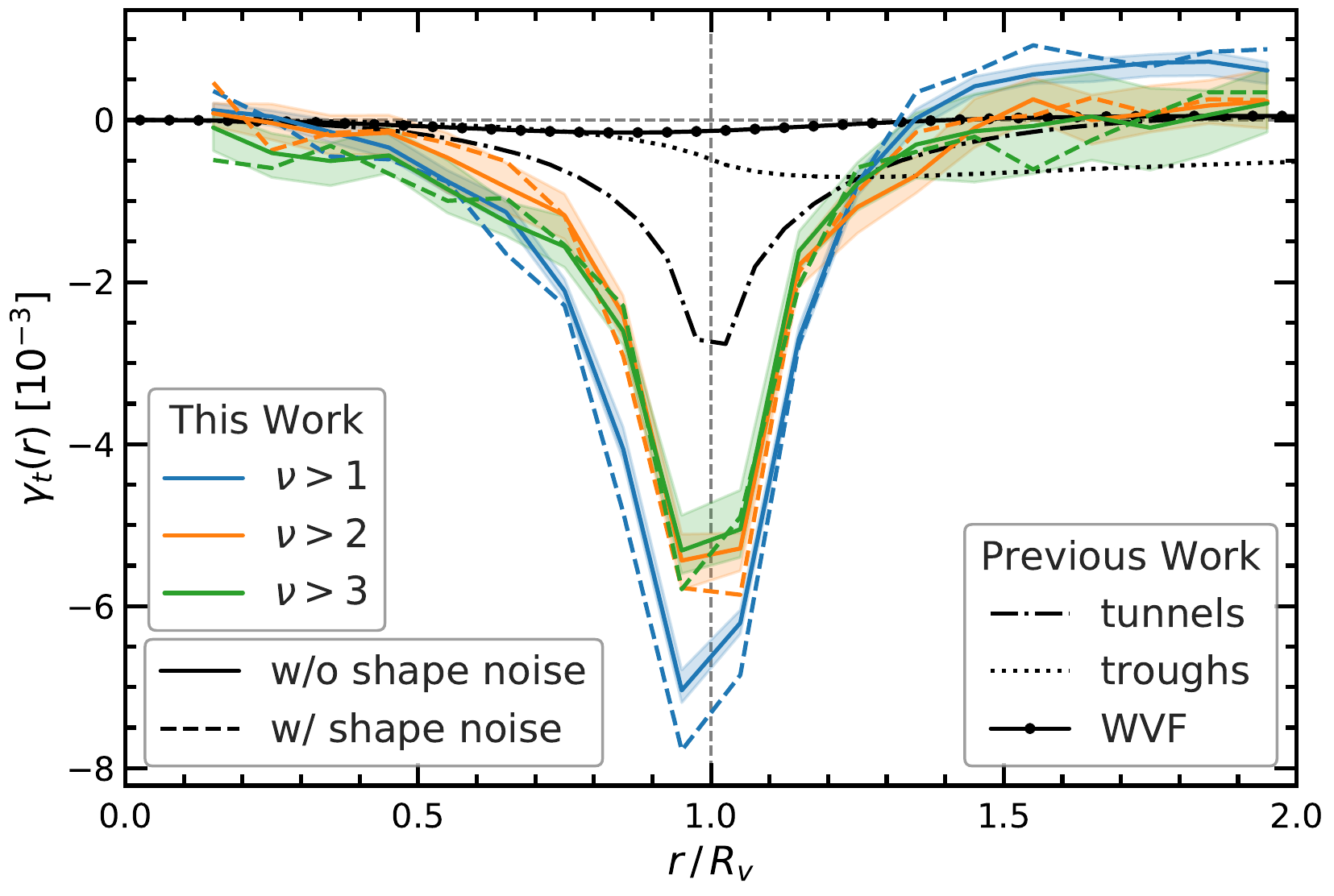}
    \vskip -.2cm
	\caption{(Colour Online) The stacked tangential shear profiles of WL voids. The coloured lines show the average profile for the void catalogues identified in this work, without GSN (solid) and with GSN (dashed). The shaded regions, show the 1$\sigma$ bootstrap uncertainties. The black patterned lines correspond to the shear profile of underdensities identified in the \textit{galaxy distribution}: tunnels (dot-dashed), troughs (dotted) and Watershed Void Finder (WVF; circle-solid) voids \citep{SHED}.
    }    
	\label{fig:Shears}
\end{figure}

Fig.~\ref{fig:Shears} shows the tangential shear profiles of the \voles{}, calculated using Eq.~\eqref{Eq: shear_from_convergence}. These profiles are qualitatively similar with the tangential shear of galaxy voids \citep[e.g.,][]{SHED} and show a maximum signal at $r=\Rv$. The maximum signal has a negative value, indicating that voids lead to diverging lensing, similar to a concave lens.
Like the convergence profiles, larger $\nu$ thresholds reduce the width associated to the maximum tangential shear signal, while the height and position of the maximum signal are almost independent of the $\nu$ threshold from $\nu > 2$. The shear signals are very similar for the case with GSN, with the greatest deviation at the lower $\nu$ threshold, $\nu > 1$ due to spurious peaks from GSN, which create spurious voids.

The error bars for the tangential shear profiles are smaller than those for the convergence profiles (see Fig.~\ref{fig:Convergences}). The uncertainty in the convergence profile is dominated by modes many times larger than the void size, which determines on average a systematic shift up or down between the convergence profiles of voids some distance apart. We checked that the voids in different regions of the lensing map have the same convergence profile up to some constant shift in $\kappa$ values. In contrast, the tangential shear profiles are only sensitive to the shape of the convergence profiles, and are not affected by a constant shift of the latter. Thus, given one WL peak catalogue, different \voles{} have very similar lensing profiles, however the profiles show a weak dependence on the $\nu$ threshold used to define the peak catalogue.
It remains to be seen whether there is a strong cosmological model dependence of the \voles{} $\gamma_t$ profiles, which we leave for future work.

Fig.~\ref{fig:Shears} also compares the \voles{} tangential shear profiles to results from other void definitions used in previous works. We compare the \voles{} with three types of voids identified in the galaxy distribution: 3D watershed voids (WVF, \citealt{Platen2007}), fixed-aperture cylinders along lines-of-sight with low projected galaxy number densities \citep[troughs;][]{Gruen:2015jhr} and tunnels (the same as the method used in this work but applied to galaxy fields). These three void catalogues are obtained from a $z=0.5$ halo occupation distribution galaxy catalogue that matches the clustering of \textsc{sdss-cmass} which has a galaxy number density, $\bar{n}_g\approx3\times10^{-4}h^3$Mpc$^{-3}$ \citep[see][for more details]{SHED}.  This represents a typical galaxy catalogue at redshift, $z=0.5$. We take the WL signals of the galaxy voids from \citet{SHED}, and we rescale them to match our source galaxies redshift, $z_s=1$.

We find that the maximum $\gamma_t$ signal of \voles{} is roughly twice as large as tunnels identified from a galaxy field, and about $10$ and $40$ times larger than the signals of troughs and WVF voids, respectively. 
The stronger lensing signal in \voles{} is not surprising since the  WL peaks are taken straight from the convergence field itself, which is directly related to the projected total matter distribution. The projected galaxy number density used to identify the \citeauthor{SHED} tunnels and troughs is ${\sim}500$ gal deg$^{-2}$ and is much larger than that used for the \voles{} catalogues (see Fig. \ref{fig:peak distribution}). We find that using higher number density peak catalogues, which could be obtained with a smaller smoothing scale and smaller GSN, results in \voles{} with an even higher tangential shear signal and thus increases further the differences between \voles{} and galaxy tunnels.

\section{Conclusions and future work}
\label{sect:cons}

Cosmic voids are becoming an increasingly important cosmological probe. While theoretical studies usually focus on voids identified from the dark matter field, in observations galaxies are usually used as tracers to find voids. Here, we have proposed an alternative: to identify voids in the WL convergence field (dubbed {\sc vole}s), which, since it represents the line-of-sight projected matter distribution, is conceptually closer to identifying voids as underdensities in the matter distribution. This opens a new window for exploring the cosmic mass distribution and, in particular, for designing novel environment-sensitive cosmological tests. 

As an example, we identified \voles{} by applying the tunnel algorithm of \citet{SHED} to the WL lensing peak distribution, and investigated several properties, including their abundance, convergence and tangential shear profiles. Using peaks with lower SNR leads to finding smaller voids, which are on average more underdense. The void convergence profiles, which are indicative of the projected matter density inside and around the voids, are negative for $r\leq0.75\Rv$, which corresponds to line-of-sight underdensities, and show a sharp overdense peak at the void edge, $r = R_v$.
In terms of tangential shear profiles, the \voles{} show a maximum signal at $r= \Rv$, with the maximum signal height somewhat independent of the peak catalogue used to identify the voids. However, the width of the tangential shear signal decreases when using peaks corresponding to a higher SNR threshold.
We found that the amplitude of the maximum signal in the tangential shear profiles for the \voles{} is roughly twice as large as that of voids generated using the same void finder (tunnels) but by using galaxies as tracers. The amplitude is more than an order of magnitude larger compared to those corresponding to galaxy voids identified using other algorithms (troughs and WVF voids). This shows the benefit of using a more reliable tracer of the projected total matter field. 

The method introduced here represents a new avenue to identify 2D voids rather than a new void-finding algorithm, in the sense that many, if not all, existing void finders (troughs, spherical void finders, WVF, {\sc zobov}, etc.) can be applied to the WL convergence maps. Indeed, in principle one can use the WL convergence field itself (i.e. not just the peaks from it) as a tracer field for void identification. The study of void identification and void lensing from the same WL observation is also convenient in practice, because there is no need for foreground galaxies, whose redshifts are hard to measure accurately and can be affected by the peculiar velocities of the galaxies. Instead, the particular lensing map that is used to study void lensing is expected to offer sufficient information for locating those very voids. With the upcoming WL surveys ({\sc hsc}, {\sc euclid}, {\sc lsst}, etc.) which offer lensing maps with increasing sky coverage, we hope that this approach will take us a step forward in extracting information from such maps in a maximal way.

Unlike galaxy voids, for \voles{} it is also possible to use void abundance to discriminate different models as there is no ambiguity in modelling the galaxy populations for these models. The universal scaled void abundance shown in Fig.~\ref{fig:cumulative void radius distribution} implies that it is possible to find generic simulation-calibrated fitting formulae for these void properties \citep[e.g.,][]{Hamaus:2014} which can be used as theoretical templates in cosmological tests. For the latter purpose, it is also critical to test the \voles{} void finding method in real weak lensing data sets to understand how observational systematics and galaxy formation physics can affect the void properties. These possibilities will be investigated in follow-up studies.

\section*{Acknowledgements}

We thank Alexandre Barreira for providing the lensing maps used in the analysis of this work, and Yanchuan Cai, Chieh-An Lin, Masato Shirasaki for useful comments. CTD is supported by a Science and Technology Facilities Council (STFC) PhD studentship through grant ST/R504725/1. MC is supported by STFC grant ST/P000541/1. BL acknowledges support by the European Research Council (ERC) through grant ERC-StG-716532-PUNCA and the STFC through grant ST/P000541/1. 
This work used the DiRAC Data Centric system at Durham University,
operated by the Institute for Computational Cosmology on behalf of the
STFC DiRAC HPC Facility (\url{www.dirac.ac.uk}). This equipment was funded
by BIS National E-infrastructure capital grant ST/K00042X/1, STFC capital
grant ST/H008519/1, and STFC DiRAC Operations grant ST/K003267/1 and
Durham University. DiRAC is part of the National E-Infrastructure.




\bibliographystyle{mnras}
\bibliography{mybib}






\bsp	
\label{lastpage}
\end{document}